\documentclass[pra,aps,amsmath,twocolumn,showpacs,superscriptaddress,
longbibliography]{revtex4-1}

\usepackage{amssymb}
\usepackage{graphicx}
\usepackage[pdftex,bookmarks,colorlinks,breaklinks]{hyperref}
\hypersetup{linkcolor=red,citecolor=blue,filecolor=dullmagenta,urlcolor=blue}
\usepackage{color}

\usepackage{soul}
\definecolor{indiagreen}{rgb}{0.07, 0.53, 0.03}

\usepackage{amsmath}
\usepackage{mathrsfs}
\usepackage{amsfonts}

\begin{document}

\title{Molecular orbitals of an elastic artificial benzene}

\author{A.~M. Mart\'inez-Arg\"uello}
\email{alael@icf.unam.mx}
\affiliation{Instituto de Ciencias F\'isicas, Universidad Nacional Aut\'onoma de M\'exico, Apartado Postal 48-3, 62210, Cuernavaca Mor., Mexico}

\author{M. P. Toledano-Marino}
\affiliation{Instituto de Ciencias F\'isicas, Universidad Nacional Aut\'onoma de M\'exico, Apartado Postal 48-3, 62210, Cuernavaca Mor., Mexico}

\author{A. E. Ter\'an-Ju\'arez}
\affiliation{Instituto de Ciencias F\'isicas, Universidad Nacional Aut\'onoma de M\'exico, Apartado Postal 48-3, 62210, Cuernavaca Mor., Mexico}

\author{E. Flores-Olmedo}
\affiliation{Departamento de Ciencias B\'asicas, Universidad Aut\'onoma Metropolitana-Azcapotzalco, Av. San Pablo 180, Col. Reynosa Tamaulipas, 02200 Ciudad de M\'exico, Mexico.}

\author{G. B\'aez}
\affiliation{Departamento de Ciencias B\'asicas, Universidad Aut\'onoma Metropolitana-Azcapotzalco, Av. San Pablo 180, Col. Reynosa Tamaulipas, 02200 Ciudad de M\'exico, Mexico.}

\author{E. Sadurn\'i}
\affiliation{Benem\'erita Universidad Aut\'onoma de Puebla, Instituto de F\'isica, Apartado Postal J-48, 72570 Puebla, M\'exico.}

\author{R.~A. M\'endez-S\'anchez}
\affiliation{Instituto de Ciencias F\'isicas, Universidad Nacional Aut\'onoma de M\'exico, Apartado Postal 48-3, 62210, Cuernavaca Mor., Mexico}


\begin{abstract}

Benzene, a hexagonal molecule with formula C$_6$H$_6$, is one of the most important aromatic hydrocarbons. Its structure arises from the $sp^{2}$ hybridization from which three in-plane $\sigma$-bonds are formed. A fourth $\pi$-orbital perpendicular to the molecular plane combines with those arising from other carbon atoms to form $\pi$-bonds, very important to describe the electronic properties of benzene. Here this $\pi$-system is emulated with elastic waves. The design and characterization of an artificial mechanical benzene molecule, composed of six resonators connected through finite phononic crystals, is reported. The latter structures trap the vibrations in the resonators and couple them through evanescent waves to neighboring resonators establishing a tight-binding regime for elastic waves. Our results show the appearance of a spectrum and wave amplitudes reminiscent of that of benzene. Finite element simulations and experimental data show excellent agreement with the H\"uckel model for benzene.

\end{abstract}


\maketitle


\section{Introduction}
\label{sec:Intro}

Current experimental techniques allow the controlled fabrication of structures at a molecular and atomic scale~\cite{Li2019,Carlotti2016,Manrique2015,Guedon2012,Champagne2005,Reichert2002,Moresco2001,Gimzewski1999}. One of the main goals in this field of research, inspired by the works of A. Aviram and M. A. Ratner~\cite{Aviram1974}, is the use of individual molecules as basic components of electronic devices~\cite{Manrique2015,Champagne2005,Liang2002,Nitzan2003,Xu2003,Kubatkin2003,Joachim2005,Tao2006,CunibertiBook,CuevasBook}. In order to do that, high control and manipulation of the electronic properties of the molecule and of the contacts it is attached to is required~\cite{Lastapis2005,Zhirnov2006,Diez2011,Geng2004,Basch2005,Haiss2006,Kronemeijer2008}. This still represents a difficult task to achieve experimentally~\cite{Carlotti2016}.

Among the several molecules showing electronic functionalities, conjugated molecules are of great interest~\cite{Li2019,Diez2011,Gholami2006,Cardamone2006,Yoshizawa2008,Andrews2008,Chen2008,Tada2015} since they possess delocalized $\pi$-electrons which are responsible for their conducting properties. Benzene has received special attention due to its size and because it has transistor functionalities~\cite{DiVentra2000}. A description and control of its molecular orbitals are of interest since the coupling of these orbitals to the outside has implications in the conducting properties of the molecule~\cite{Diez2011,Tada2015,Walczak2004,Kornilovitch2001}. Consequently, to get a further insight into the conducting and electronic properties of $\pi$-systems, some theoretical models have been used~\cite{CunibertiBook,CuevasBook}.

A paradigmatic model widely used for describing solid state systems and the electronic properties of $\pi$-molecular structures is the tight-binding (TB) model~\cite{CunibertiBook,CuevasBook}.
It takes into account interference effects, retains atomic and electronic structure details, and also captures important features like circular currents~\cite{Andrews2008,Mujica1994,Kostyrko2002,Walczak2004,Rai2010,Yuta2014,Cattena2010,Nozaki2010,Nozaki2012,Nozaki2017,Rix2019}. 
In this model the electrons are imagined to be strongly bonded to atoms and interact weakly with neighboring atoms. 
Thus the electronic wave function of a system of interacting atoms is written in terms of the wave functions of a system of isolated atoms, plus the coupling that is usually considered to nearest neighbors. This model has been recently emulated in classical wave systems with the use of optical fibers~\cite{Kondakci2015} and in microwave experiments using dielectric resonators~\cite{Barkhofen2013}, the latter playing the role of atomic sites, from which artificial analogues of graphene~\cite{Bellec2013} and polyacetylene have been constructed~\cite{Stegmann2017}. The TB model can also be emulated with mechanical systems. An elastic resonator, whose wave amplitude is localized in a certain region in space and with an exponential decay out of it, is required for its implementation. This localized wave constitutes an \emph{artificial atomic orbital}~\cite{Filiberto2020}. Additionally, to establish the nearest-neighbor tight-binding regime for elastic waves, at least two such orbitals must be evanescently coupled to each other. Opposite to the optical and microwave cases the mechanical TB model allows a high nearest-neighbor coupling control, which is often the most dominant part; next-nearest neighbor interactions can also be easily engineered. Here the design and characterization of a mechanical artificial benzene molecule, in which the analogues of the spectrum and molecular orbitals of benzene are emulated, is reported. Our experimental results are in agreement with finite-element numerical simulations of the artificial molecule and with the simplest tight-binding model for benzene.

The paper is organized as follows. In the next section, we discuss the design of the relevant artificial elastic atomic orbital. The benzene-like molecule in the TB regime for elastic waves is shown in section~\ref{sec:Artificial-Benzene}. The experimental characterization of the artificial benzene is explained in section~\ref{sec:Experiment}. Finally, we present our conclusions in section~\ref{sec:Conclusions}.

\begin{figure*}
\centering
\includegraphics[width=15.0cm]{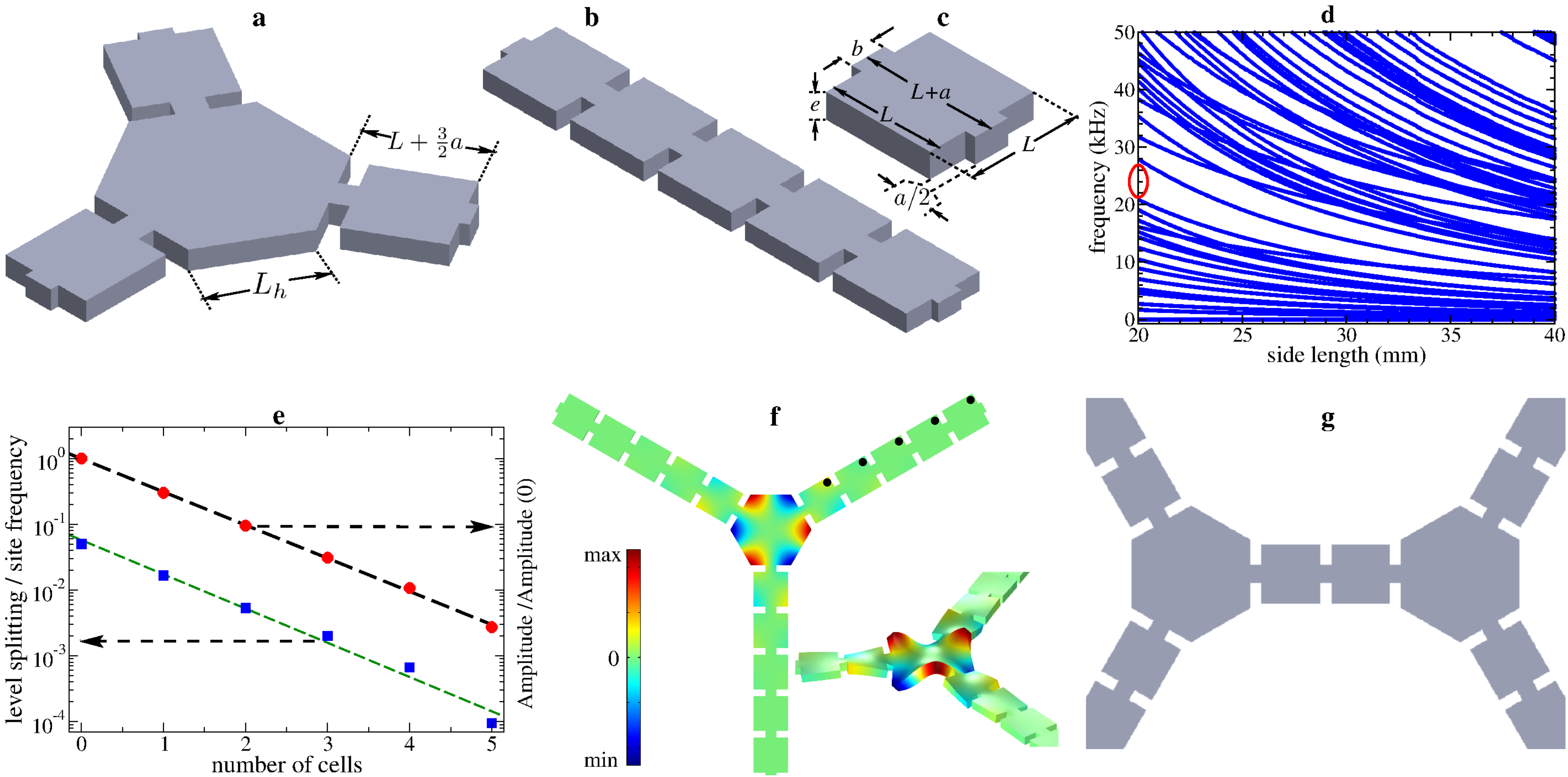}
\caption{{\footnotesize (Color online) \textbf{a} Elastic artificial atomic site composed of a hexagonal-shaped resonator of length $L_{h}$ and connected to three bonds with an angle of $120^{\circ}$ between them. \textbf{b} Finite phononic crystal (FPC), formed by the repetition of a unit cell, that plays the role of bonds. \textbf{c} The unit cell is composed of a square-shaped aluminum plate of thick $e$ and side $L$, and two smaller plates, one on each side of the larger plate, of size $a/2$ and length $b$. The FPCs give rise to bands and gaps in its level dynamics (frequency spectrum as a function of $L$), as shown in panel~\textbf{d} for a structure composed of five unit cells. The bandgap is indicated by a red oval. By choosing a frequency of the resonator in the gap of the FPCs, the vibration is trapped in the resonator as shown in panels~\textbf{e} and~\textbf{f}, the maximum (minimum) amplitude in panel~\textbf{f} is shown in red (blue) color. In the lower-right part, this vibration is shown in 3D. An exponential decay of the amplitude of vibration along the bonds, in the positions indicated in black-filled circles, is also evident. This exponentially decaying wave amplitude, panel~\textbf{e} in red-filled circles, emulates a quantum atomic orbital and is used to build the artificial benzene molecule. The amplitude is measured at the points indicated in black-filled circles in figure~\ref{fig:FigureOne} \textbf{f}.  An exponential decay of the frequency level splitting as a function of the number of FPC cells (see panel \textbf{g}) is obtained, as shown in panel \textbf{e} in blue-filled squares. The dashed lines are fits to the corresponding data. The geometrical parameters are $e=6.05$~mm, $L=20.0$~mm, $a/2=2.5$~mm, $b=8.2$~mm, and $L_{h} = 30.4$~mm.}} 
\label{fig:FigureOne}
\end{figure*}
%


%
\begin{figure}
\centering
\includegraphics[width=8.5cm]{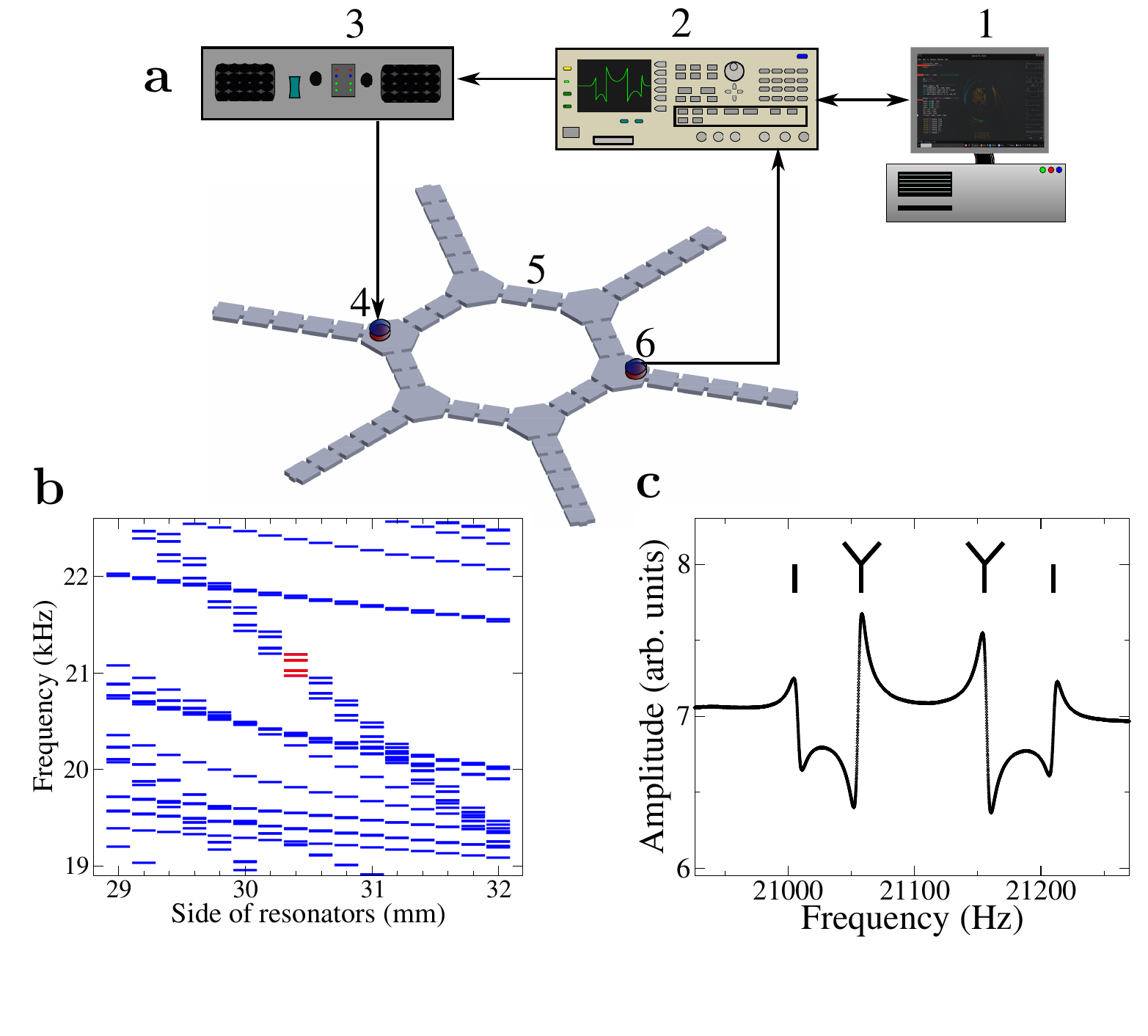}
\caption{{\footnotesize (Color online) \textbf{a} Experimental setup. 1 Workstation, 2 Vector network analyzer Anritsu MS4630B, 3 Cerwin Vega CV-2800 high-fidelity audio amplifier, 4 electromagnetic acoustic transducer (EMAT) exciter, 5 elastic artificial benzene, 6 EMAT detector. \textbf{b} Numerical frequency spectrum of artificial benzene as a function of $L_h$, the FPCs give rise to forbidden bands that trap the vibrations in the resonators. For $L_h = 30.4$ mm, in the interval from 20500 to 21500 Hz, some natural frequencies associated with the resonators (red horizontal lines) are observed within the forbidden band. \textbf{c} Experimental frequency spectrum, six resonances are observed: two singlets and two doublets, the vertical lines correspond to the numerical predictions. An excellent agreement is observed between the experimental and numerical results. }}
\label{fig:FigureTwo}
\end{figure}

\section{Artificial atomic orbital}
\label{sec:Artificial-Atom}

In order to establish the tight-binding regime for mechanical waves, a structure whose wave amplitude is spatially localized in a certain region and with an exponential decay out of it, is required. Furthermore, this structure has to be able to connect with another structure to form an artificial molecule. The localized wave amplitude or \emph{artificial elastic atomic orbital} has to couple in an evanescent way with another artificial atomic orbital to form \emph{artificial elastic molecular orbitals}. A first 1D example of such elastic artificial molecule was reported in~\cite{Filiberto2020}. For its implementation, a mechanical resonator was connected to finite phononic crystals (FPCs), playing the role of bonds, which trap the resonator vibrational wave amplitude in its position and with an exponential decay along the bonds, emulating in this way an atomic orbital. Furthermore, allowing such mechanical resonators to weakly interact with each other via the FPCs, a 1D elastic crystal, which is in complete agreement with a quantum tight-binding model~\cite{Filiberto2020}, was constructed. Following this line, we use the artificial atomic site shown in Fig.~\ref{fig:FigureOne} \textbf{a}. It is composed of a hexagonal-shaped two-dimensional elastic resonator of side length $L_{h}$ connected to three FPCs (bonds), see Figs.~\ref{fig:FigureOne} \textbf{b} and \textbf{c}. Both, the resonator and FPCs have the same thickness $e$. 
The angle between bonds is $120^{\circ}$. This shape allows a configuration that is typical in molecules with alternating single and double chemical bonds that yield to an $sp^{2}$ hybridization, and are found in planar molecules like ethylene and benzene, among others. 
The FPCs are formed by the repetition of a unit cell consisting of a square-shaped plate of length $L$ and two small plates of side $a/2$ and length $b$, one at each side of the larger plate (see Fig.~\ref{fig:FigureOne} \textbf{c}). The FPCs give rise to allowed and forbidden bands in its frequency spectrum, as shown in Fig.~\ref{fig:FigureOne}~\textbf{d} for a structure composed of five unit cells. 
By choosing a vibrating frequency $F$ of the resonator in the bandgap of the FPC, the latter traps the vibration in the resonator and produces an exponential decay of the amplitude along the bonds as can be observed in Figs.~\ref{fig:FigureOne}~\textbf{e} and~\textbf{f}. 
The evanescence length $\xi \approx 0.86$, in units of the maximum amplitude, 
is in perfect correspondence with the fact that a localized state decays exponentially by means of an imaginary wave number $ 
\mathrm{e}^{-\kappa \hat{\kappa} \cdot \vec{x} }$ with $\kappa = 1/\xi$. 
This trapped vibrational wave amplitude is in agreement with the case of electrons strongly bonded to their atoms and emulates a single quantum atomic orbital~\cite{Filiberto2020}. 
The structure mentioned above is used to build the artificial elastic benzene.
In figure~\ref{fig:FigureOne} \textbf{e} the level splitting for two coupled artificial atomic sites as function of the cells of the coupler is also given. The coupling between neighboring artificial sites is given by half the level splitting ($\nu = 56$~Hz).


\section{Artificial elastic benzene}
\label{sec:Artificial-Benzene}

The benzene molecule is a hydrocarbon composed of six carbon atoms, bonded to one hydrogen atom each, and bonded between them by means of an $sp^{2}$ hybridization. In this hybridization the relevant orbitals, concerning the wave conduction, are the $\pi$-orbitals. These orbitals can be described within the tight-binding approximation from which an analogous molecule can be constructed in a mechanical structure. For this purpose, we couple six identical artificial atomic sites (see figure~\ref{fig:FigureOne}~\textbf{a}) in a hexagonal structure to form the artificial benzene molecule. 
The six $\pi$ atomic orbitals, ignoring the hydrogen atoms, are emulated assuming that the hexagonal structure comes from the $sp^{2}$ hybridization of the carbon atoms and to the bonding to the hydrogen atoms. This mechanical analogue reproduces the expected structure of the frequency spectrum, with the repulsion between the eigenfrequencies of the resonators, and also the molecular orbitals, or wave amplitudes, that are reminiscent to that of the benzene molecule. 
In figure~\ref{fig:FigureTwo} \textbf{a} we show the artificial elastic benzene molecule. 
The geometrical parameters are the same as those of the artificial atomic site of figure~\ref{fig:FigureOne}~\textbf{a} and are determined by means of finite element simulations using the software COMSOL Multiphysics. 
FPCs composed of five unit cells are taken to the outside to avoid edge effects. FPCs with only two unit cells connect the resonators.
This guarantees that the eigenfrequencies belonging to the resonators are in a bandgap of the FPCs in a frequency range from $19$ to $22$~kHz, as shown in figure~\ref{fig:FigureTwo} \textbf{b} in red horizontal lines. 
As expected, six eigenfrequencies show up: two singlets at higher and lower frequencies and two doublets in between. The structure of the eigenfrequency spectrum and the corresponding eigenmodes associated to each eigenfrequency are in excellent agreement with the ones obtained from the tight-binding model for benzene, as shown below. 

The numerical frequency spectrum for the artificial molecule is
\begin{eqnarray}
f_{1} & = & 20989\, \mathrm{Hz} , \nonumber \\
f_{2} & = & 21042\, \mathrm{Hz} , \quad f_{2'} = 21043\, \mathrm{Hz} , \nonumber \\
f_{3} & = & 21153\, \mathrm{Hz} , \quad f_{3'} = 21155\, \mathrm{Hz} , \quad \mathrm{and} \nonumber \\
f_{4} & = & 21209\, \mathrm{Hz}
\end{eqnarray}
where we observe that $f_{2} \approx f_{2'}$ and $f_{3} \approx f_{3'}$, i.e., the modes of vibration, or artificial elastic molecular orbitals, associated to those eigenfrequencies are degenerate, and represent doublets in the molecule. 
Thus, the artificial system shows six frequency levels, two of them degenerate with a frequency difference $\Delta f \approx 53$ Hz between singlets and doublets and $2\Delta f$ ($\approx 111$ Hz) between doublet and doublet. 
This structure shown by the frequency levels is observed in the energy spectrum of the benzene molecule in the quantum mechanical case. 
Furthermore, the modes of vibration associated to those eigenfrequencies also reproduce the symmetry of the molecular orbitals of benzene.
In the nearest-neighbor tight-binding approximation for the artificial benzene the following Hamiltonian is assumed
\begin{equation}
\label{eq:Huckel}
H = \left(
\begin{array}{cccccc}
F & -\nu & 0 & 0 & 0 & -\nu \\
-\nu & F & -\nu & 0 & 0 & 0 \\  
0 & -\nu & F & -\nu & 0 & 0 \\  
0 & 0 & -\nu & F & -\nu & 0 \\  
0 & 0 & 0 & -\nu & F & -\nu \\  
-\nu & 0 & 0 & 0 & -\nu & F 
\end{array}
\right),
\end{equation}
where all {\em site frequencies}, analog to the site energies, are equal to $F=21099$~Hz. 
The {\em hopping frequencies}, analog to the hopping energies, are all $\nu=56$~Hz. 
The wave amplitudes $\psi_{n}(\vec{r})$ of interacting artificial atoms are expanded in terms of those of isolated artificial atoms, $\phi_{m}(\vec{r})$ as
\begin{equation}
\label{eq:Expansion}
\psi_{n}(\vec{r}) = \sum_{m} c_{nm} \phi_{m}(\vec{r}),
\end{equation}
where $c_{nm}$ are the coefficients in the expansion. 
More details of the tight-binding approximation, and of the solutions, can be found in Appendix~\ref{sec:AppendixA} and in Ref.~\cite{Hernandez2020}.

\begin{figure*}
\centering
\includegraphics[width=16.0cm]{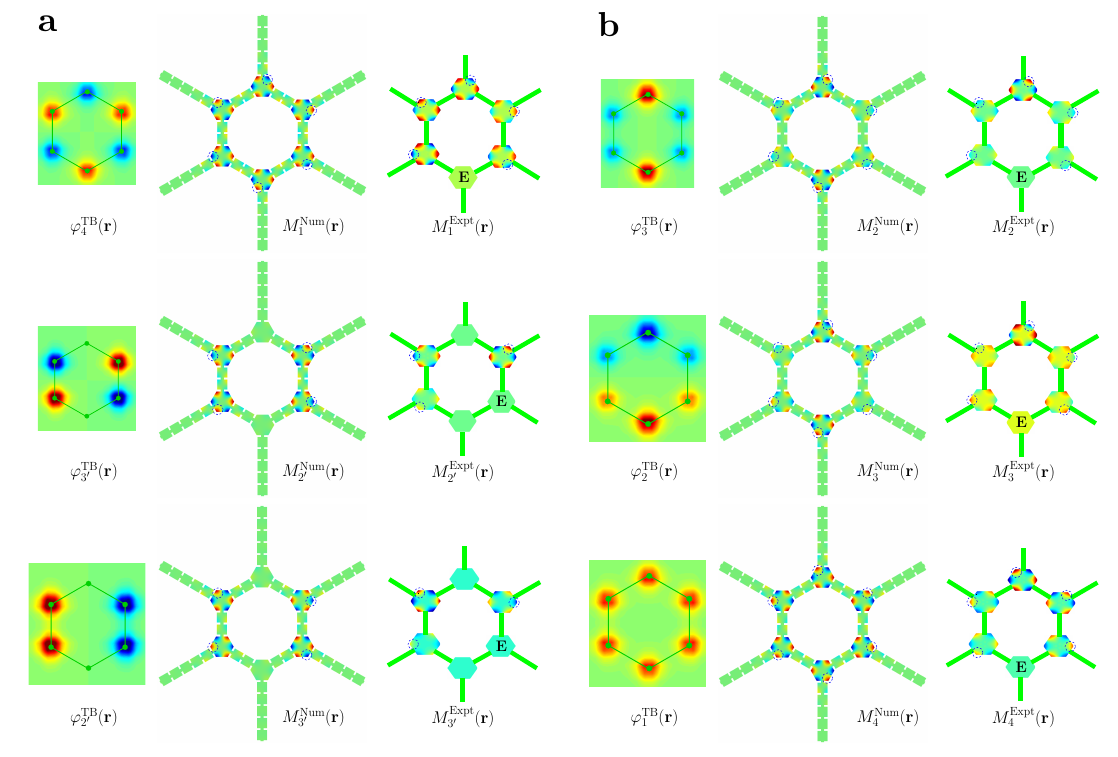}
\caption{{\footnotesize (Color online) Wave amplitudes from tight-binding model, artificial mechanical benzene, and experiments, in left, middle and right columns of panels~\textbf{a} and~\textbf{b}. The color scale is the same of Fig.~\ref{fig:FigureOne}. The blue-dashed circles in the center and right columns of~\textbf{a} and~\textbf{b} indicate equivalent points and help to identify the symmetry of the artificial orbital. The ``E'' in the experimental orbitals indicate the  position of the EMAT exciter.}}
\label{fig:TNE_orbitals}
\end{figure*}

Figure~\ref{fig:TNE_orbitals} shows the molecular orbitals obtained from the tight-binding model for benzene of (\ref{eq:Huckel}), in left columns of panels \textbf{a} and \textbf{b}. An excellent approximation to both the structure of the frequency spectrum and the symmetries of the molecular orbitals of the benzene molecule is observed~\cite{CuevasBook}. Furthermore, we note that the modes of vibration obtained from the numerical artificial molecule also reproduce the symmetries of the molecular orbitals of benzene, as shown in the same panels of figure~\ref{fig:TNE_orbitals} in the middle columns. In the molecular orbitals, $\varphi_{j}(x, y)$, each atomic site of the molecule in red (blue) color corresponds to equivalent points in the artificial benzene. These points are indicated in blue-dashed circles in figure~\ref{fig:TNE_orbitals}, panels \textbf{a} and \textbf{b}. 
As can be observed, the mechanical artificial orbitals $M_{j}(\textbf{r})$ reproduce the symmetries of  the molecular orbitals $\varphi_{j}(x,y)$. 
These results show the successful emulation of the underlying properties of the benzene molecule like its spectrum structure and its molecular orbitals.


\section{Experiment}
\label{sec:Experiment}

The predictions obtained from the numerical simulations for the artificial benzene can be tested experimentally with the setup of Figure~\ref{fig:FigureTwo}~\textbf{a}. The artificial benzene molecule, with geometrical parameters given in Fig.~\ref{fig:FigureOne}, machined from a single aluminum plate (Young modulus $E = 71.1$~GPa, mass density $\rho = 2.708\mathrm{kg}/ \mathrm{m}^3$, and Poisson's ratio $\nu = 0.36$), was held up by nylon threads attached to support stands. The threads are tensioned by the weight of the structure and maintain minimal contact with the system avoiding an alteration of the vibrations of the artificial molecule.

The setup consists of a workstation that controls and extracts the data from a vector network analyzer (VNA) Anritsu MS4630B. 
The VNA generates a harmonic signal of frequency $f_0$ which is sent to a Cerwin-Vega CV-2800 high-fidelity audio amplifier. 
The intensified signal is sent to an electromagnetic acoustic transducer (EMAT), consisting of a coil and a permanent magnet~\cite{Morales2001}. 
The EMAT was located close to a vertex of one resonator of the artificial molecule, without physical contact. 
The former excites out-of-plane waves when the dipole moment axis of both the coil and permanent magnet coincide with the normal to the plate. 
A second EMAT, of smaller dimensions, with the same configuration of coil and permanent magnet as the exciter, located close to the farthest resonator, detects the response that is sent back to the VNA. The data are analyzed in the workstation. 

To obtain the artificial molecular orbitals we do the following procedure. Firstly, a set of measurements on a grid of a sixth of one resonator is performed. Secondly, due to the symmetry of the mode in the resonator (see figure~\ref{fig:FigureOne}~\textbf{f}), this sixth is reflected to build the full artificial atomic orbital. Finally, to obtain the artificial molecular orbital in any resonator a measure of its amplitude in equivalent points is multiplied by the atomic orbital, previously obtained. This gives a total of four modes each one associated to its corresponding eigenfrequency. 
The other two modes, associated to the degeneracies, are obtained as before, but the vibrations in two opposite resonators are suppressed by clamps. 
These modes of vibration correspond to the artificial benzene molecular orbitals and reproduce the symmetry of those of benzene obtained from the TB model and other more sophisticated models like density-functional theory~\cite{CuevasBook}.

Figure~\ref{fig:FigureTwo} \textbf{c} shows the measured frequency spectrum. As expected six resonances show up, two of them degenerate: two singlets and two doublets. This is the analogous spectrum to the energy spectrum of the benzene molecule. An asymmetry of these resonances, due to the Fano effect~\cite{Angel2015}, is observed. In the right columns of panels \textbf{a} and \textbf{b} of figure~\ref{fig:TNE_orbitals} the measured modes are shown. These are the analogous orbitals reminiscent to the orbitals of the benzene molecule.


\section{Conclusions}
\label{sec:Conclusions}

We have presented a design and experimental realization of a benzene-like elastic artificial molecule. The analogous spectrum and $\pi$-orbitals which are reminiscent to that of benzene were emulated. The experimental frequency spectrum and wave amplitudes agree with the theoretical predictions from finite element simulations and the simplest tight-binding model for benzene. 

In the literature, artificial analogues have been used to get a further understanding of the wave behavior of molecular systems with the advantage of an easy manipulation of the systems. They also provide experimental realizations to verify theoretical predictions that are difficult to attain at quantum level. 
Furthermore, elastic systems in particular are of interest since it has been found that the quality factor in these systems is sometimes orders of magnitude higher than those found in microwave systems. The successful emulation of $\pi$-orbitals and the tight-binding regime for elastic waves opens the door to study more complicated molecules and of paradigmatic models of solid state physics. In addition, our results provide high control over the nearest-neighbor and of the orbitals, which are of interest since the coupling of these orbital to the outside contacts has strong implications in the conducting properties of molecules. Here, by manipulating the geometrical parameters of the resonators or bondings allows to control the coupling strength of the amplitudes of vibrations as required.


\acknowledgments

This work was supported by CONACYT and DGAPA-UNAM under projects CB-2016/184096 and PAPIIT IN111021, respectively.
AMM-A acknowledges financial support from DGAPA-UNAM through the POSDOC program.
GB was supported by CONACYT under project CB2017-2018/AI-S-33920.


\appendix
\addcontentsline{toc}{chapter}{Apendices}

\section{Nearest-neighbor tight-binding approximation for benzene}
\label{sec:AppendixA}

In the TB model, the frontier electrons of atoms are assumed to be well localized around the ions. 
The electronic wave functions of interacting atoms are expanded in terms of the wave functions of isolated atoms, or individual atomic orbitals. 
That is,
\begin{equation}
\label{eq:Expansion2}
\psi_{n}(\vec{r}) = \sum_{m} c_{nm} \phi_{m}(\vec{r}),
\end{equation}
where $c_{nm}$ are the coefficients of the expansion in the base of atomic orbitals $\{ \phi_{m} \}$. This is in essence the method known as linear combination of atomic orbitals, or LCAO. 

For molecules like benzene formed by atoms of the same species, the nearest-neighbor tight-binding model consists in taking equal energies, $\varepsilon$, for each atomic site, and that the  electrons can only ``hop" between nearest neighbors with a hopping energy $\nu$. 
Then, the TB model for benzene can be written as~\cite{CuevasBook}
\begin{equation}
\label{eq:Huckel2}
H = \left(
\begin{array}{cccccc}
\varepsilon & -\nu & 0 & 0 & 0 & -\nu \\
-\nu & \varepsilon & -\nu & 0 & 0 & 0 \\  
0 & -\nu & \varepsilon & -\nu & 0 & 0 \\  
0 & 0 & -\nu & \varepsilon & -\nu & 0 \\  
0 & 0 & 0 & -\nu & \varepsilon & -\nu \\  
-\nu & 0 & 0 & 0 & -\nu & \varepsilon 
\end{array}
\right),
\end{equation}
where we note that the matrix element $H_{16} = H_{61}$, which is due to the cyclicity of the molecule. This tight-binding model for benzene is completely equivalent to the so-called \emph{H\"uckel approximation}, which considers only the atomic $\pi$ orbitals since the electrons in the $\sigma$ orbitals are strongly bonded to the atom and they have very weak interactions to one another.

The diagonalization of matrix~(\ref{eq:Huckel2}) yields the eigenenergies
\begin{eqnarray}
\label{eq:Eigenvalores}
E_{1} & = & \varepsilon - 2\nu ,\nonumber \\
E_{2} = E_{2'} & = & \varepsilon - \nu , \nonumber \\
E_{3} = E_{3'} & = &  \varepsilon + \nu , \nonumber \\
E_{4} & = & \varepsilon + 2\nu ,
\end{eqnarray}
which consist of six levels: two singlets, $E_{1}$ and $E_{4}$, and two doubly degenerate levels or doublets, $E_{2} = E_{2'}$ and $E_{3} = E_{3'}$. These degeneracies are due to the ring structure of the molecule as well as time-reversal symmetry~\cite{Hernandez2020}. The energy difference between singlets and doublets, and between doublet and doublet, is $\Delta E = -\nu$ and $2\Delta E=-2\nu$, respectively. This structure of the energy spectrum is a characteristic feature of benzene. A common way of writing the eigenstates is~\cite{CuevasBook}
{\small{
\begin{eqnarray}
\label{eq:Eigenestados}
|\varphi_{1} \rangle & = & \frac{1}{\sqrt{6}} (| \phi_{1} \rangle + | \phi_{2} \rangle + | \phi_{3} \rangle + | \phi_{4} \rangle + | \phi_{5} \rangle + | \phi_{6} \rangle) , \nonumber \\ 
|\varphi_{2} \rangle & = & \frac{1}{\sqrt{12}} (2 | \phi_{1} \rangle + | \phi_{2} \rangle - | \phi_{3} \rangle - 2 | \phi_{4} \rangle - | \phi_{5} \rangle + | \phi_{6} \rangle) , \nonumber \\
|\varphi_{2'} \rangle & = & \frac{1}{2} (| \phi_{2} \rangle + | \phi_{3} \rangle - | \phi_{5} \rangle - | \phi_{6} \rangle) , \nonumber \\
|\varphi_{3} \rangle & = & \frac{1}{\sqrt{12}} (2 | \phi_{1} \rangle - | \phi_{2} \rangle - | \phi_{3} \rangle + 2 | \phi_{4} \rangle - | \phi_{5} \rangle - | \phi_{6} \rangle) , \nonumber \\
|\varphi_{3'} \rangle & = & \frac{1}{2} (| \phi_{2} \rangle - | \phi_{3} \rangle + | \phi_{5} \rangle - | \phi_{6} \rangle ) , \nonumber \\
|\varphi_{4} \rangle & = & \frac{1}{\sqrt{6}} (| \phi_{1} \rangle - | \phi_{2} \rangle + | \phi_{3} \rangle - | \phi_{4} \rangle + | \phi_{5} \rangle - | \phi_{6} \rangle).
\end{eqnarray}
}}
Each molecular orbital, $|\varphi_{j} \rangle$ ($j=1,2, 2', 3, 3',$ and 4), is a combination of the atomic orbitals ($\pi$-orbitals), $|\phi_{n}\rangle$ ($n=1, 2, \ldots, 6$), and can be expressed in configuration space, where a reasonable assumption is to consider that the $\pi$ atomic orbitals decay exponentially out of its position, with an evanescence length $\xi$. Under this assumption the atomic orbitals take the form
\begin{equation}
\phi_{j}(r) = A\, \mathrm{exp} \Big(-\frac{|x_{j} - x|}{\xi} - \frac{|y_{j} - y|}{\xi} \Big),
\end{equation}
where $(x_{j}, y_{j})$ indicate the position coordinates of the atomic site $j$ and $A$ is a normalization constant. Considering the molecular structure of benzene, with a distance between sites of $1.4$ \AA, and only the $\pi$ orbitals, a reasonable estimate to the decay length leads to $\xi \approx 0.176$ \AA~\cite{Hernandez2020}.



\end{document}